\documentclass[fleqn,12pt,twoside]{article}

\usepackage[headings]{espcrc1}

\usepackage{graphicx}
\usepackage{subfigure}
\usepackage{epsfig}
\usepackage{amssymb}
\usepackage{calrsfs}
\usepackage{mathrsfs}
\usepackage{graphics}
\usepackage{pslatex}

\newcommand{\sqrtsnn}{\sqrt{s_{\mbox{\tiny{\it{NN}}}}}}
\newcommand{\sqrts}{\sqrt{s}}
\def\mean#1{\ensuremath{\left<#1\right>}}

\newcommand\qhat{{\mean{\hat{q}}}}

\newcommand{\dNdeta}{dN_{ch}/d\eta|_{\eta=0}}

\title{High-energy heavy-ions physics: from RHIC to LHC}

\author{David d'Enterria\address[CERN]{CERN, PH-EP, CH-1211 Geneva 23, Switzerland}}
       
\runtitle{High-energy heavy-ions physics: from RHIC to LHC}
\runauthor{D. d'Enterria}

\begin{document}

\maketitle

\begin{abstract}
\noindent
A selection of experimental results in high-energy nucleus-nucleus collisions after 
five years of operation of the Relativistic Heavy-Ion Collider (RHIC) is presented. 
Emphasis is put on measurements that provide direct information on fundamental 
properties of high-density QCD matter. The new experimental opportunities accessible 
at LHC are introduced, in particular those that may help clarify some of the current open 
issues at RHIC.
\end{abstract}


\section{Introduction}

Quantum Chromodynamics (QCD) is the only quantum field theory of the Standard Model whose
{\it collective} behaviour (phase diagram and phase transitions) is accessible to study in the 
laboratory. High-energy heavy-ion (AA) collisions offer the only experimental means known 
so far to concentrate a significant amount of energy ($\mathscr{O}$(1,10 TeV) at RHIC,LHC)
in a ``large'' volume ($\mathscr{O}(100$ fm$^3$) at thermalization times of $\tau_0\approx$ 1 fm/$c$).
The study of the many-body dynamics of high-density QCD covers a vast range of fundamental 
physics problems:

\begin{itemize}
\item {\bf Deconfinement and chiral symmetry restoration}:
Lattice QCD calculations predict a new form of matter at energies densities above 
$\varepsilon\approx$ 1 GeV/fm$^3$ consisting of an extended volume of deconfined and bare-mass 
quarks and gluons: the Quark Gluon Plasma (QGP)~\cite{latt}. The scrutiny of this new 
state of matter (equation-of-state, order of the phase transition, ...) promises to shed light on 
fundamental questions of the strong interaction such as the nature of confinement, the mechanism 
of mass generation (chiral symmetry breaking, structure of the QCD vacuum) and hadronization, which 
still evade a thorough theoretical description due to their highly non-perturbative nature~\cite{millenium_prizes}.

\item {\bf Early universe cosmology}: The quark-hadron phase transition took place some 10 $\mu$s 
after the Big-Bang and was the most important event taking place in the Universe between the electro-weak 
(or SUSY) transition ($\tau\sim 10^{-10}$ s) and Big Bang nucleosynthesis ($\tau\sim$ 200 s). 
Depending on the order of the QCD phase transition\footnote{The order itself is not exactly known: 
the pure SU(3) gauge theory is first-order whereas introduction of 2+1 flavours makes it of a fast 
cross-over type~\cite{latt}.}, several cosmological implications such as the formation of strangelets 
and cold dark-matter (WIMP) clumps or baryon fluctuations leading to inhomogeneous nucleosynthesis, 
have been postulated~\cite{Schwarz:2003du}.

\item {\bf Parton structure and evolution at small-$x$}: At high energies, hadrons consist of a very 
dense system of gluons with small (Bjorken) parton fractional momenta $x=p_{parton}/p_{hadron}$. 
At low-$x$, the probability to emit an extra gluon is large $\propto\alpha_s\ln(1/x)$ and non-linear 
$gg$ processes will eventually dominate the parton evolution in the hadronic wave functions. 
Whereas HERA results indicate that for $x\gtrsim$10$^{-3}$, the parton evolution with $Q^2$ (or $\ln(1/x)$) 
is described by the usual DGLAP (or BFKL) equations, at lower values of $x$ and around a saturation momentum of 
$Q^2_s \sim$ 2 GeV$^2$, such a configuration is theoretically described in terms of the 
``Colour Glass Condensate'' (CGC) picture~\cite{iancu03}. Since the nonlinear growth of the gluon 
density depends on the transverse size of the system (i.e. $Q_s^2\propto A^{1/3}$, where $A$ is the number 
of nucleons in the nucleus), the effects of gluon saturation are expected to set in earlier 
for ultrarelativistic heavy nuclei than free nucleons.

\item  {\bf Gauge/String duality}: Theoretical calculations based on the Anti-de-Sitter/Conformal-Field-Theory 
(AdS/CFT) correspondence permit to obtain results in strongly coupled 
($\lambda = g^2\,N_c\gg 1$) SU($N_c$) gauge theories in terms of a weakly-coupled dual gravity 
theory~\cite{ads_cft}. Recent applications of this formalism for QCD-like ${\cal N}=4$ super Yang-Mills 
theories have allowed to compute  transport properties of experimental relevance, - such as the QGP 
viscosity~\cite{kovtun04}, the ``jet quenching'' parameter $\qhat$~\cite{wiedem06}, 
or the heavy-quark diffusion coefficient~\cite{casalderey_teaney06} -, 
from black hole thermodynamics calculations. 
Such results provide valuable insights on dynamical properties of non-perturbative QCD that cannot be 
directly treated by 
lattice methods.
\end{itemize}

In this overview, we present a {\it selection} of  experimental results (mostly from the comprehensive 
reviews of the 4 RHIC experiments~\cite{phenix_wp,star_wp,phobos_wp,brahms_wp})  from AuAu, dAu 
and pp collisions up to a maximum center-of-mass energy of $\sqrtsnn$ = 200 GeV. Direct information 
on the thermodynamical and transport properties of the strongly interacting medium produced in AA 
collisions is commonly obtained by comparing the results for a given observable $\Phi_{AA}$ to those measured 
in  p(d)A (``cold QCD matter'') and in pp (``QCD vacuum'') collisions as a function of center-of-mass energy, 
$p_T$, rapidity $y$,  reaction centrality (impact parameter $b$), and particle type (mass): 
\begin{equation}
R_{AA}(\sqrtsnn,p_T,y,m;b) \,=\,\frac{\mbox{\small{``hot/dense QCD medium''}}}{\mbox{\small{``QCD vacuum''}}}\,
\propto \,\frac{\Phi_{AA}(\sqrtsnn,p_T,y,m;b)}{\Phi_{pp}(\sqrts,p_T,y,m)}
\end{equation}
The observed {\it enhancements}  (e.g. in photon or baryon yields, or soft hadron slopes) and/or 
{\it suppressions} (e.g. in total multiplicities, high-$p_T$ leading hadrons, or quarkonia yields) in the 
$R_{AA}(\sqrtsnn,p_T,y,m;b)$ ratios can be directly related to the properties of the produced QCD matter after 
accounting for a realistic modeling of the space-time evolution of the collision process\footnote{The hot 
and dense systems produced in heavy-ion collisions at RHIC expand longitudinally (transversely) with 
$\mean{\beta}\approx 1.0(0.6)$ and stop self-interacting collectively at freeze-out times $\tau\approx$ 15 fm/$c$.}.


\section{Reduced hadron multiplicities $\mapsto$ Saturated gluon distribution function $x\,G_{A}(x,Q^2)$ ?}

\begin{figure}[htb]
\begin{center}
\includegraphics[width=7.5cm,height=7cm]{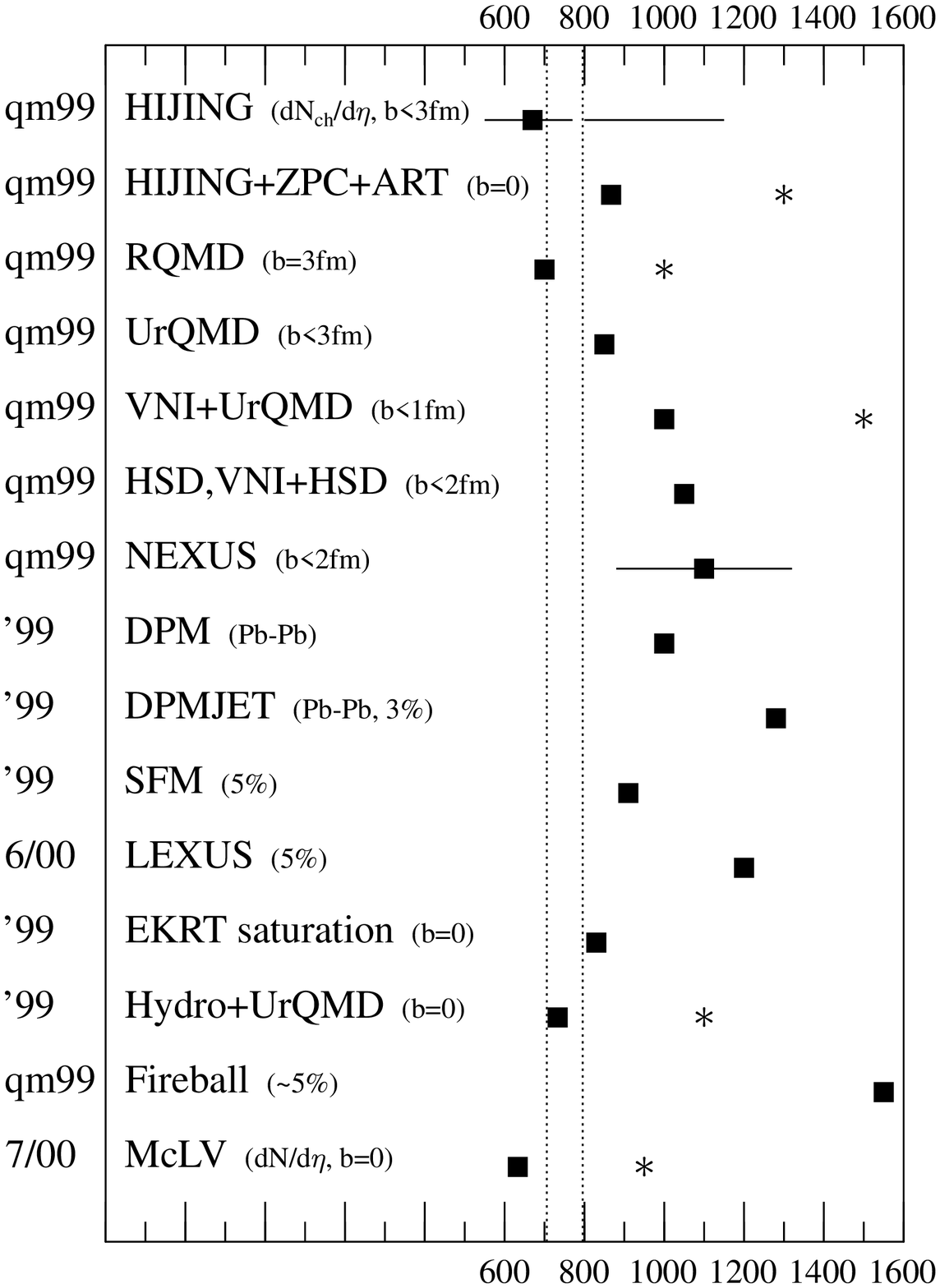}
\includegraphics[width=8.4cm]{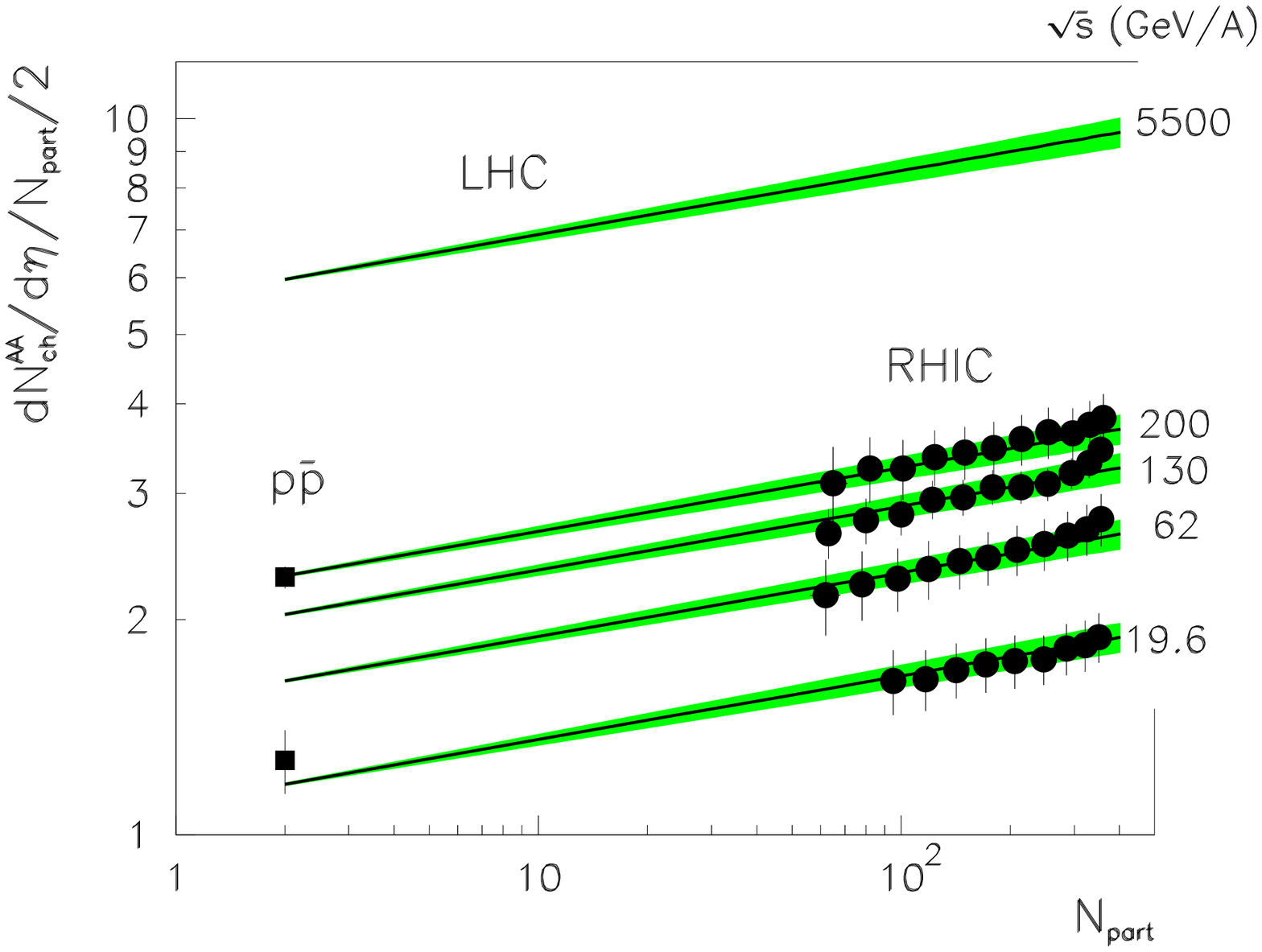}
\vskip -0.7cm
\caption{Left: Data versus models for $\dNdeta$ in AuAu at $\sqrtsnn$ = 200 GeV~\cite{eskola_qm01}.
Right: Energy and centrality dependences (in terms of the number of nucleons participating in the collision, 
$N_{\rm part}$) of  $\dNdeta$  (normalized by $N_{\rm part}$):
PHOBOS AuAu data~\protect\cite{phobos_wp} versus the predictions of the saturation approach~\protect\cite{armesto04} . }
\label{fig:dNdeta}
\end{center}
\end{figure}

The bulk hadron multiplicities measured at mid-rapidity in central AuAu at $\sqrtsnn$ = 200 GeV are 
$\dNdeta\approx$ 700, comparatively lower than the $\dNdeta\approx$ 1000 expectations~\cite{eskola_qm01} 
of ``minijet'' dominated scenarios, soft Regge models (without accounting for strong 
shadowing effects), or extrapolations from an incoherent sum of proton-proton collisions
(Fig.~\ref{fig:dNdeta}, left). On the other hand, approaches~\cite{kharzeev,armesto04} based on gluon 
saturation~\cite{iancu03}, which take into account a reduced {\it initial} number of 
scattering centers in the nuclear parton distribution functions, $f_{a/A}(x,Q^2)<A\cdot f_{a/N}(x,Q^2)$, 
agree well with experimental data. In those CGC calculations, the final hadron multiplicities are assumed 
to be simply related to the initial number of released partons (local parton-hadron duality) which are 
depleted in the initial state compared to proton-proton collisions due to non-linear gluon-gluon fusion effects. 

The good reproduction of the bulk AA hadron multiplicities (including its centrality and center-of-mass 
energy dependences, Fig.~\ref{fig:dNdeta} right) has been one of the supporting arguments in favour of the 
existence of non-linear QCD effects in high-energy nuclear collisions. In addition, the BRAHMS 
observation of suppressed yields of moderately high-$p_T$ hadrons in dAu at forward rapidities 
($\eta\approx$ 3.2)~\cite{brahms_wp}, as well as the ``geometrical scaling''-like behaviour 
of the nuclear PDFs observed for $x<$0.017~\cite{armesto04}, are also consistent with gluon saturation
expectations. It is worth noting, however, that both results are in a kinematic range with relatively low
momentum scales,  $\mathscr{O}$(1 GeV), where non-perturbative effects can blur a simple interpretation 
based on partonic degrees of freedom alone. Indeed, at RHIC (and HERA) energies the saturation momentum, 
- the scale at which non-linear effects 
become important and start to saturate the parton densities -, is in the transition between the
soft and hard regimes ($Q_s^2\approx$ 2 GeV$^2$). At LHC the relevance of low-$x$ QCD effects 
in hadronic collisions will be certainly enhanced due to the increased (i) center-of-mass energy $\sqrtsnn$, 
(ii) nuclear radius $A^{1/3}$, and (iii) rapidity $y$ of the produced partons.  Indeed, at LHC energies
not only the relevant Bjorken $x$ values 
will be 30--70 times lower than at RHIC: 
$x_{2}^{min} = (p_T/\sqrtsnn)e^{-y}\approx 10^{-3}(10^{-5})$ at central (forward) rapidities 
for processes with a hard scale $p_T\sim$10 GeV, but also the saturation momentum,
$Q_s^2 \sim A^{1/3}  s_{\mbox{\tiny{\it{NN}}}}^{\lambda/2}\approx$ 5 -- 10 GeV$^2$~\cite{kharzeev},
will be in the perturbative range.


\section{Strong radial and elliptic collective flows $\mapsto$ QGP as a perfect fluid ?}

\begin{figure}[htb]
\begin{center}
\includegraphics[width=7cm]{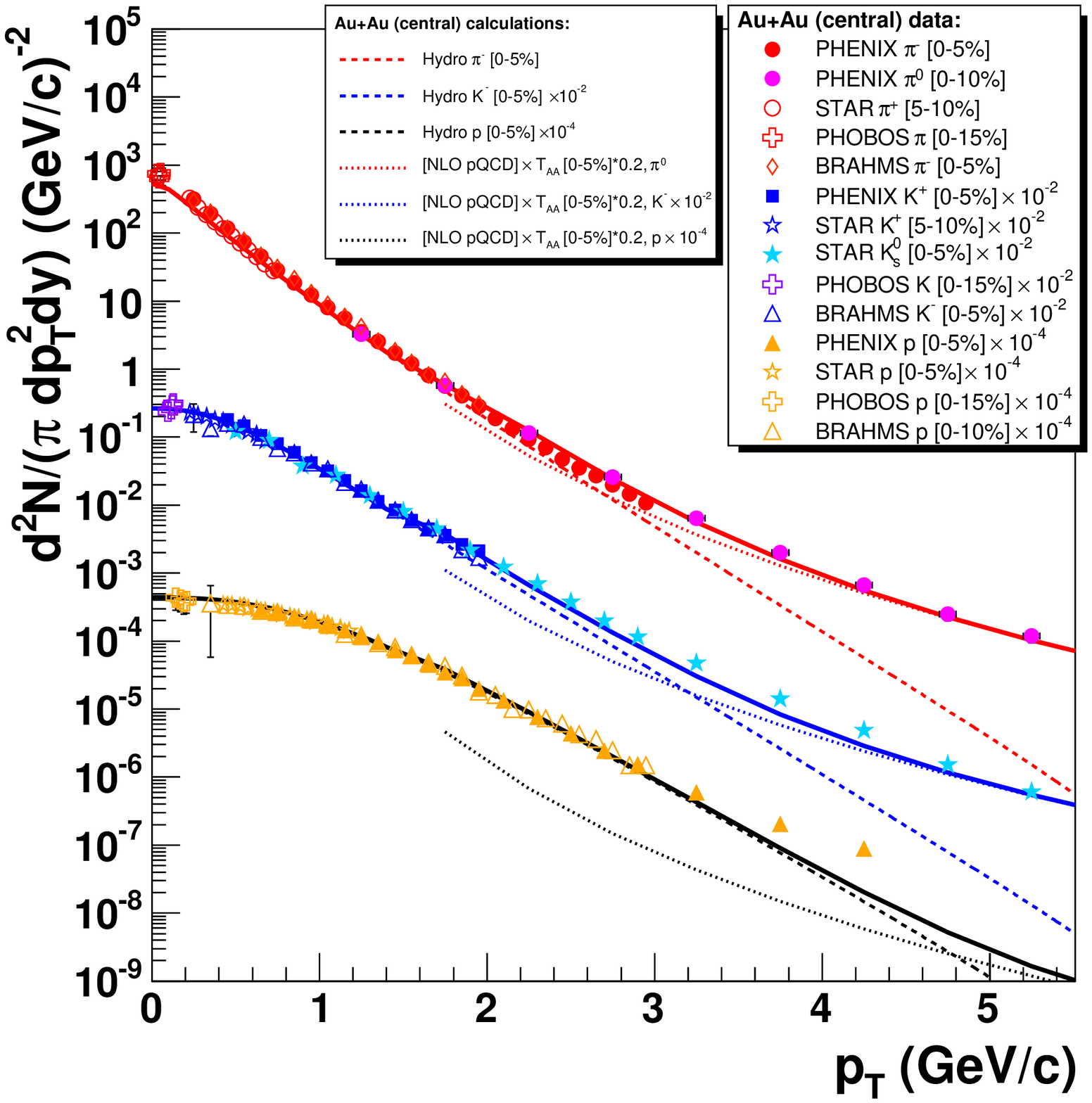}
\includegraphics[width=8cm]{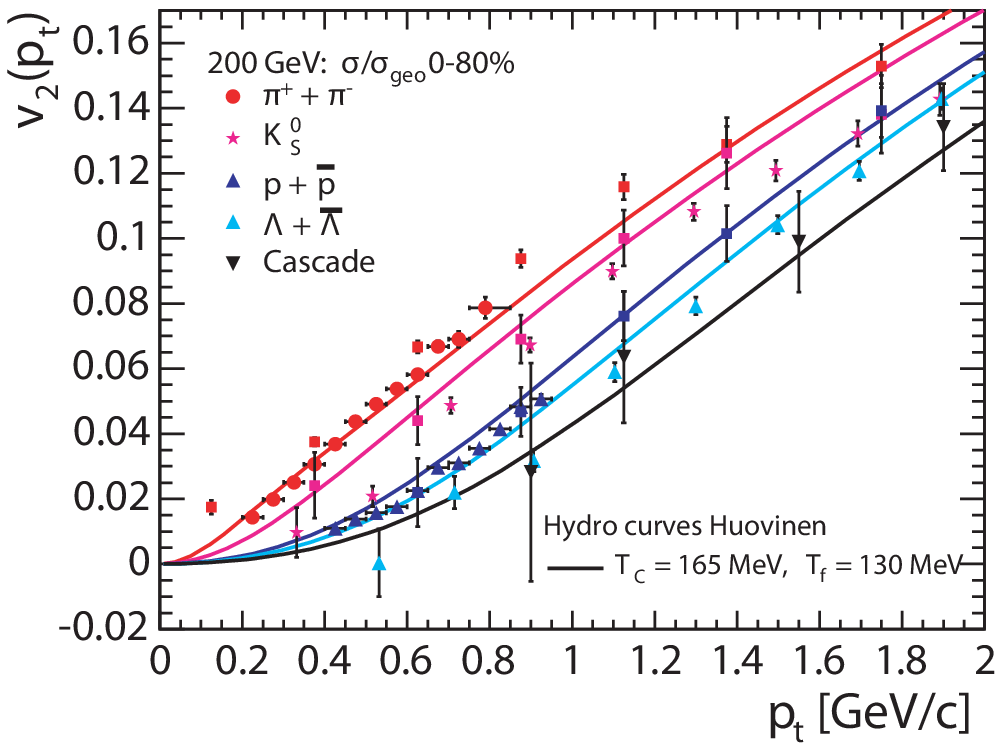}
\vskip -0.7cm
\caption{Left:  Transverse spectra for pions, kaons, and protons measured below $p_T\approx$ 5 GeV/$c$ 
in 0-10\% most central AuAu collisions at $\sqrtsnn$ = 200 GeV, compared to hydrodynamics(+pQCD)
calculations~\protect\cite{dde_peressou}. Right:  Measured elliptic flow parameter $v_2(p_T)$ for a variety of 
hadrons~\protect\cite{snellings05} compared to hydrodynamic predictions~\protect\cite{huovinen_ruuskanen06}.}
\label{fig:hydro}
\end{center}
\end{figure}

The bulk of hadron production ($p_T\lesssim$ 2 GeV/$c$) in AuAu at RHIC shows strong collective 
effects known as radial and elliptic flows. On the one hand, the measured single hadron $p_T$
spectra have an inverse slope parameter larger than that measured in pp collisions, increasing
with reaction centrality and hadron mass as expected if collective expansion effects blue-shift the 
hadron spectra (Fig.~\ref{fig:hydro}, left). Phenomenological fits to ``blast wave''  models yield transverse flow 
velocities $\mean{\beta}\approx$ 0.6~\cite{phenix_wp}. On the other hand, the azimuthal distribution 
$dN/d\phi$ of hadrons emitted w.r.t. the reaction plane show a strong harmonic modulation with a 
preferential ``in-plane'' emission in non-central collisions. Such an azimuthal flow pattern is a 
truly collective effect (absent in pp collisions) consistent with an efficient translation of the initial $x$-space 
anisotropy in non-central reactions (with an almond-shape overlap zone) into a final ``elliptical'' asymmetry 
in $p$-space. The amount of elliptic flow is quantified via the second Fourier coefficient $v_2=\mean{cos(2\phi)}_{p_T}$ 
of the $dN/d\phi$ distribution relative to the reaction plane. The large $v_2\approx$ 0.2 measured 
in the data (Fig.~\ref{fig:hydro}, right) indicates a strong degree of collectivity (pressure gradients) developing 
in the first instants of the collision. Indeed,  elliptic flow develops in the initial phase of the reaction and quickly 
self-quenches beyond $\tau\approx$ 5 fm/$c$ as the original spatial eccentricity disappears. This is confirmed 
by the observation that not only light hadrons  but charm quarks (indirectly measured via the semileptonic 
decays of D mesons into $e^\pm$) show a $v_2$ signal as large as 10\%~\cite{akiba05} clearly consistent 
with strong collective correlations during the {\it partonic} phase of the reaction. 

Interestingly, the strong $v_2$ seen in the data is inconsistent with the much lower values, $v_2\lesssim$ 6\%, 
expected for purely hadronic matter~\cite{UrQMD} as well as for a partonic system cascading with 
perturbative cross-sections ($\sigma_{gg}\approx$ 3 mb)~\cite{molnar01}. The magnitude, $p_T$ and mass 
dependences of the radial and elliptic  flows below $p_T\approx$ 2 GeV/$c$ are, however, remarkably well 
described by {\it ideal} hydrodynamics models whose space-time evolution starts with a QGP equation-of-state 
(EoS) with initial $\varepsilon_0\approx$ 30 GeV/fm$^3$ at very short thermalization times 
$\tau_0\approx$ 0.6 fm/$c$~\cite{kolb_heinz_rep,teaney_hydro,hirano,huovinen_ruuskanen06} (Fig.~\ref{fig:hydro}).  
The fast (local) thermalization times, the robust 
collective flow generated in the first instants of the reaction, and the excellent agreement of the data 
with ideal hydrodynamics models which assume a fluid evolution with zero viscosity, 
have been presented~\cite{shuryak03,thoma04,hirano05,peshier05,heinz05} as evidences that the QGP 
formed at RHIC is  a strongly coupled liquid (characterized by Coulomb coupling parameter
$\Gamma = \mean{E_{pot}}/\mean{E_{kin}}>$ 1)  rather than a weakly interacting gas of partons.
Estimates of the maximum amount of viscosity allowed by the $v_2(p_T)$ data~\cite{hirano05} give a 
value close to the conjectured universal lower bound for the dimensionless viscosity/entropy ratio, 
$\eta/s=\hbar/(4\pi)$ , obtained from AdS/CFT calculations~\cite{kovtun04}. 
This result would make the QGP the most perfect fluid ever observed. The measurement of the 
differential elliptic flow properties in AA collisions at LHC will be of primary importance to confirm such 
an interpretation and search for a possible transition from a hotter weakly interacting QGP to the liquid-like 
state found at RHIC~\cite{thoma04,hirano05}.


\section{High $p_T$ hadron suppression $\mapsto$ dense QGP with $dN^g/dy\sim$ 1000, $\qhat\sim$ 14 GeV$^2$/fm}

\begin{figure}[htb]
\begin{center}
\includegraphics[width=8.5cm,height=5.5cm]{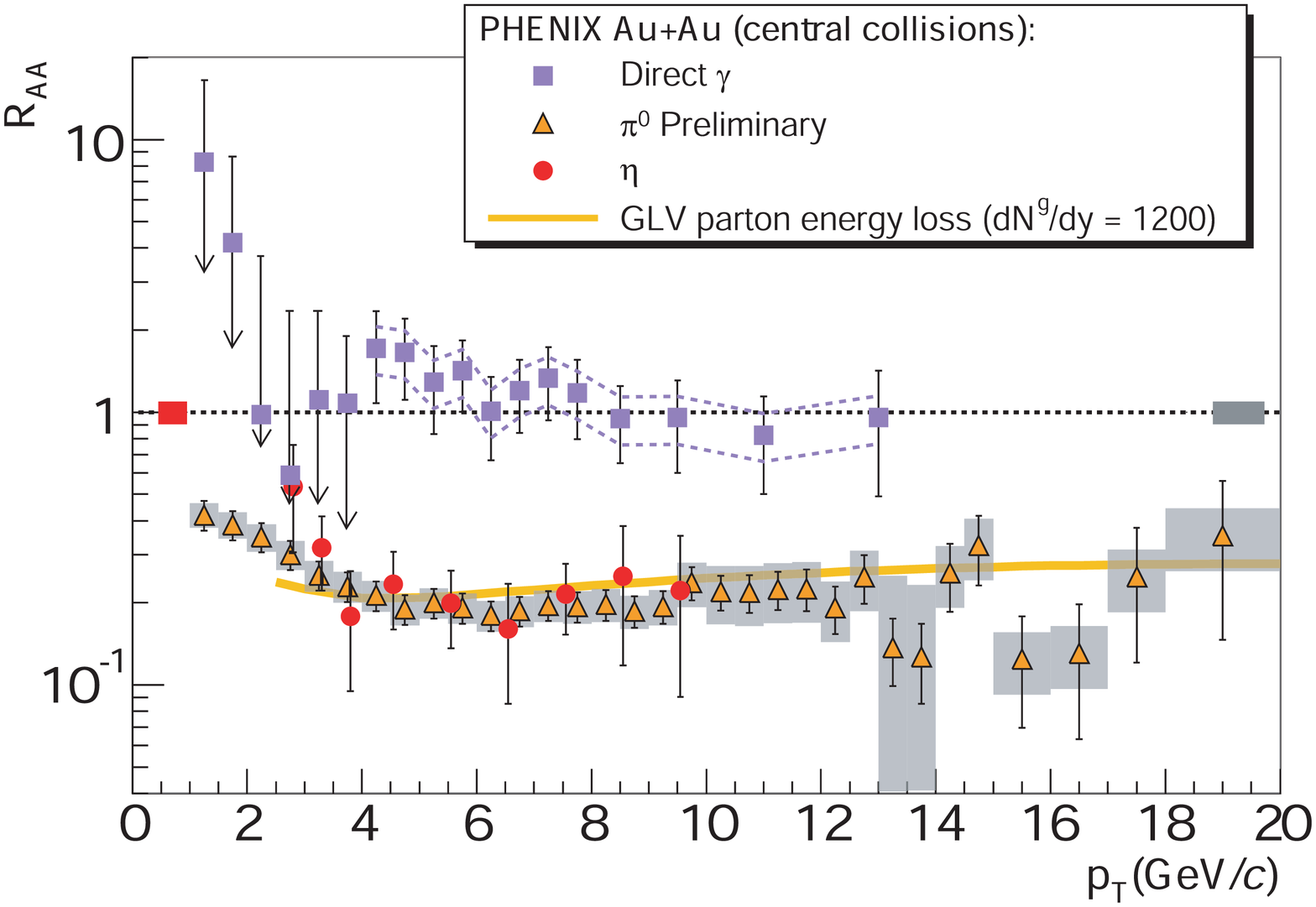}
\includegraphics[width=7.cm,height=5.25cm,clip=true]{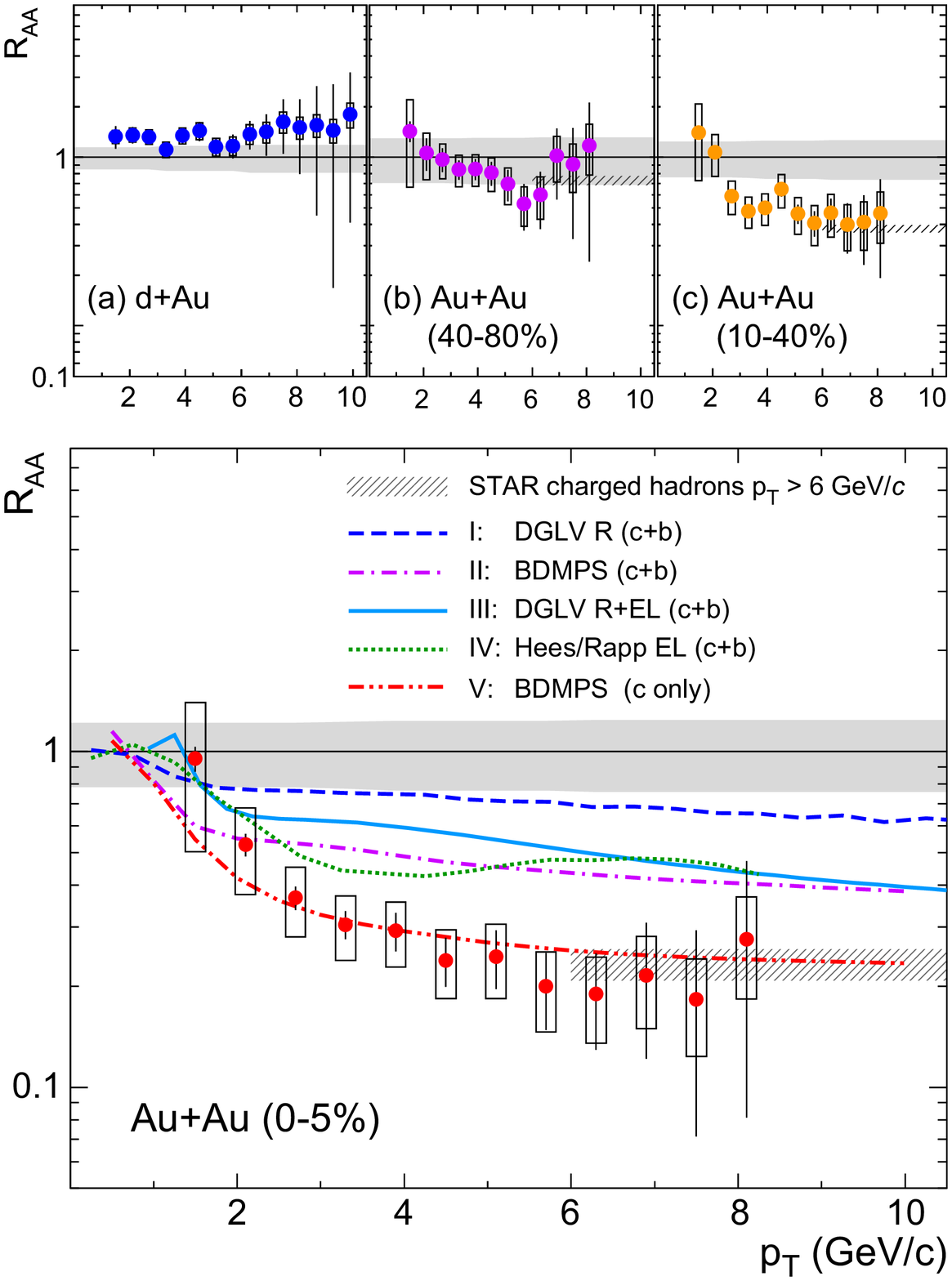}
\vskip -0.7cm
\caption{Nuclear modification factor $R_{AA}(p_T)$ for different particles measured 
in central AuAu at $\sqrtsnn$ = 200 GeV compared to parton energy loss model predictions.
Left: Direct $\gamma$, $\pi^0$, and $\eta$~\protect\cite{phenix_hipt_pi0_eta_AuAu200}. 
Right: ``Non-photonic'' $e^\pm$, mainly from D,B meson decays~\protect\cite{star_hipt_nonphoton_elect_AuAu200}.}
\label{fig:R_AA_RHIC_200}
\end{center}
\end{figure}

Among the most exciting results of the RHIC physics programme is the observed strong suppression of 
high-$p_T$ leading hadron spectra in central AA~\cite{rhic_hipt} consistent with the predicted energy loss of
the parent light quarks and gluons traversing the dense colored medium (``jet quenching'')~\cite{gyulassy90}. 
Above $p_T\approx$ 5 GeV/$c$, $\pi^0$, $\eta$, and inclusive charged hadrons show all a common factor of 
$\sim$5 suppression compared to an incoherent superposition of pp collisions 
(Fig.~\ref{fig:R_AA_RHIC_200}, left)~\cite{phenix_hipt_pi0_eta_AuAu200}. 
The $R_{AA}$ = 1 perturbative expectation which holds for other hard probes such as ``color blind'' direct 
photons and for high-$p_T$ hadrons in dAu reactions (where no final-state dense and hot system is produced)~\cite{rhic_hipt_dAu},
is badly broken ($R_{AA}\approx$ 0.2) in central AuAu collisions. 
Most of the empirical properties of the quenching factor for light-flavor hadrons (magnitude, $p_T$-, centrality-,
$\sqrtsnn$- dependences of the suppression)~\cite{dde_hp04} are in quantitative agreement with the predictions 
of non-Abelian parton energy loss models which assume that the parent parton loses energy by gluonstrahlung 
while traversing a medium with a large color density. Very large initial gluon rapidity densities, 
$dN^g/dy\approx$ 1000~\cite{vitev_gyulassy}, or equivalently, transport coefficients\footnote{$\mean{\hat{q}}$ 
characterizes the squared average $k_T$ transfer from the medium to the hard parton per unit distance~\cite{bdmps}.}, 
$\mean{\hat{q}}\approx$ 14 GeV$^2$/fm~\cite{dainese04,armesto05}, are needed to explain the amount of 
hadron suppression. However, the fact that the quenching factor for high-$p_T$  electrons from semi-leptonic 
$D$ and $B$ decays is as suppressed as the light hadrons in central AuAu (Fig.~\ref{fig:R_AA_RHIC_200}, 
right)~\cite{phenix_hipt_nonphoton_elect_AuAu200,star_hipt_nonphoton_elect_AuAu200} is in apparent conflict 
with the robust $\Delta E_{Q} < \Delta E_{q} <  \Delta E_{g}$ prediction of radiative energy loss models. In order to 
reproduce the high $p_T$ open charm/bottom suppression, jet quenching models require either initial gluon densities 
($dN^g/dy\approx$ 3000) inconsistent with those needed to describe the quenched light hadron spectra~\cite{djordj04,wicks05}, 
or a smaller relative contribution of $B$ relative to $D$ mesons than theoretically expected in the measured decay 
electron $p_T$ range~\cite{armesto05}. This discrepancy may point to an additional contribution from elastic 
(i.e. non-radiative) energy loss~\cite{mustafa03,peshier06} for heavy-quarks~\cite{wicks05} which was considered 
negligible so far. The unique possibility at LHC of fully reconstructing $c,b$ jets will be very valuable to clarify
the response of strongly interacting matter to fast {\it heavy} quarks, and provide detailed additional information
on the transport properties of QCD matter~\cite{casalderey_teaney06}.


\section{Modified semihard di-hadron $\phi$ correlations $\mapsto$ QGP speed of sound $c_s$ ?}

\begin{figure}[htbp]
\includegraphics[width=7.5cm,height=7.cm]{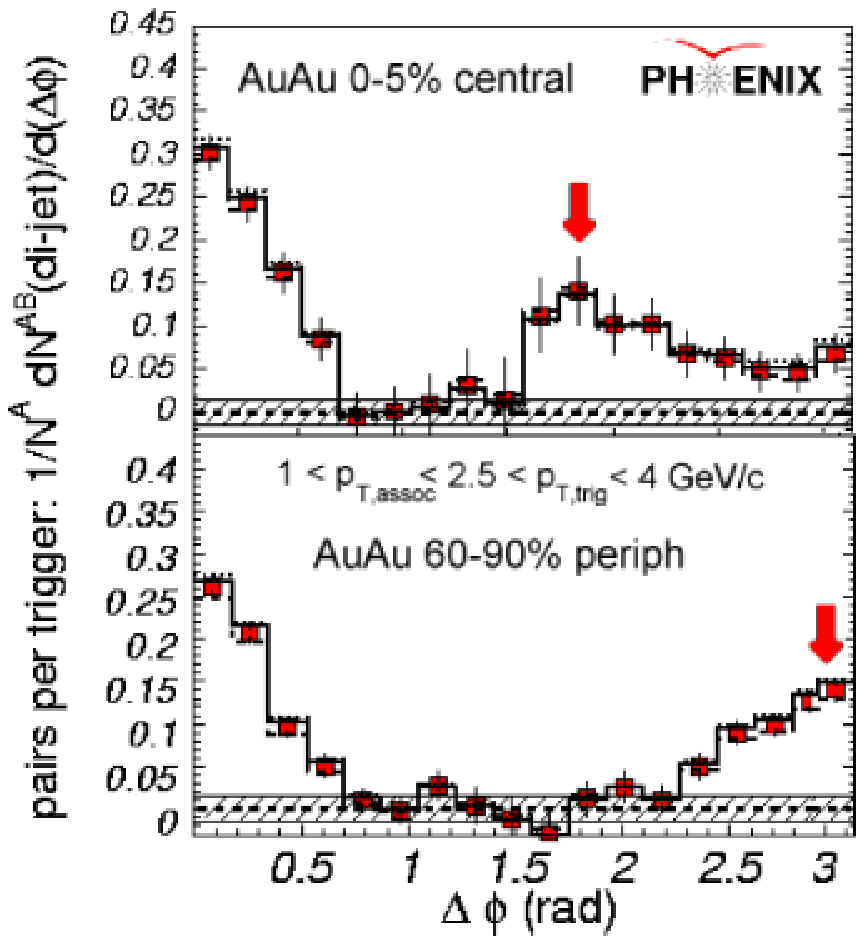}
\hskip 0.5cm
\includegraphics[width=7.8cm,height=6.8cm]{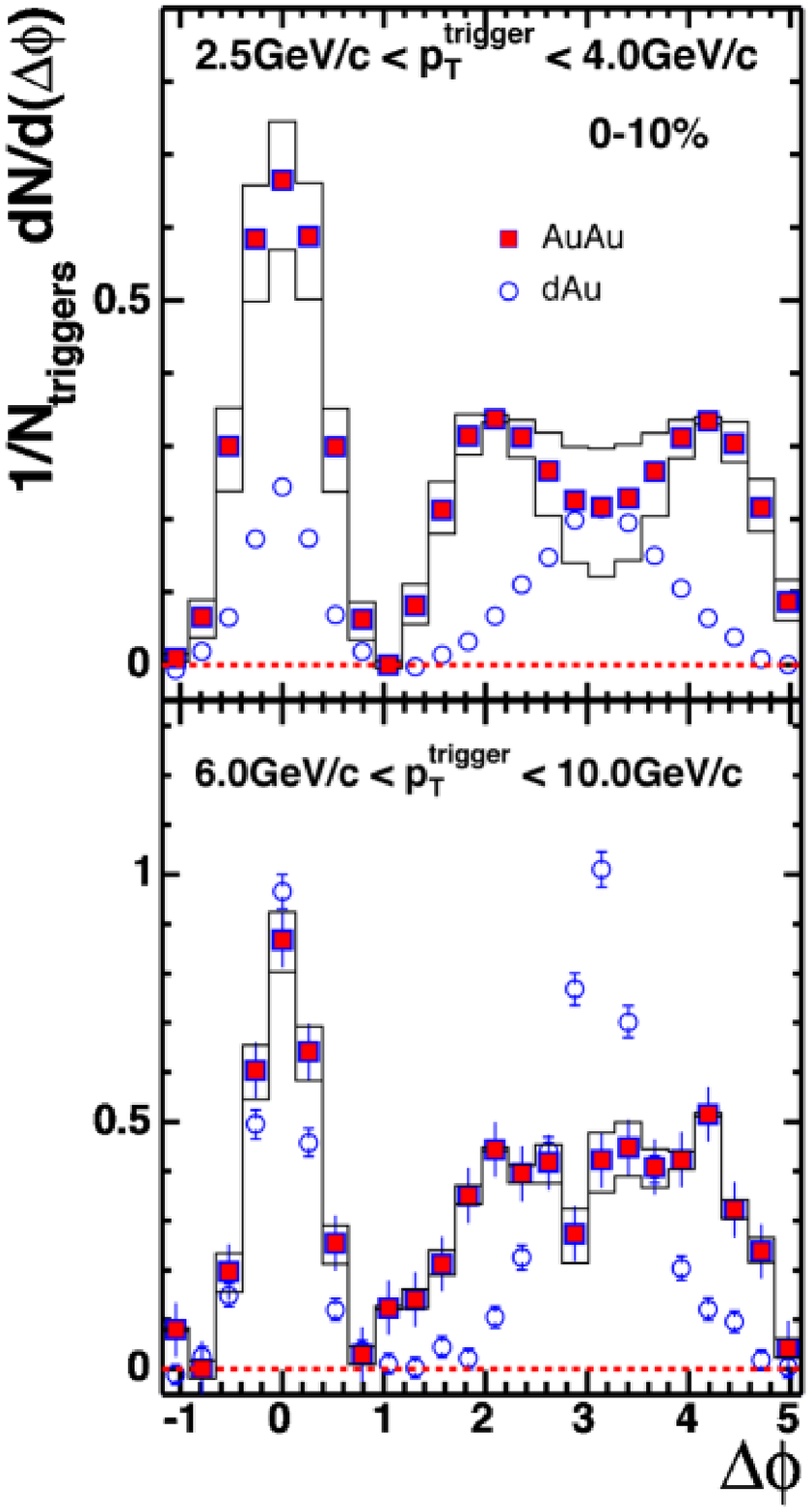}
\vskip -0.7cm
\caption{Azimuthal distributions of semihard hadrons ($p_{T,\,assoc}$ = 1 -- 2.5 GeV/$c$) 
with respect to a trigger hadron measured at RHIC. 
Left: PHENIX data in central (top) and peripheral (bottom) AuAu~\protect\cite{phenix_machcone}. 
Right: STAR data in central AuAu and dAu collisions~\protect\cite{star_machcone}.}
\label{fig:dNdphi}
\end{figure}

A second striking observation of the jet-quenching phenomena at RHIC is the strongly modified azimuthal 
dijet correlations compared to baseline pp results.  Due to the difficulties of full jet reconstruction
in AA at RHIC, jet-like correlations are measured on a statistical basis by selecting the highest $p_T$ 
{\it trigger} hadron of the event and measuring the azimuthal ($\Delta\phi = \phi - \phi_{trig}$)
and rapidity ($\Delta\eta = \eta - \eta_{trig}$) distributions of {\it associated} hadrons ($p_{T,\,assoc}<p_{T,\,trig}$):
$d^2N_{pair}/d\Delta\phi d\Delta\eta$. In pp collisions, a dijet signal appears 
as two distinct back-to-back Gaussian peaks at $\Delta\phi\approx$  0 
(near-side) and at $\Delta\phi\approx\pi$ (away-side). At variance with this standard dijet topology in the 
QCD vacuum, the away-side dihadron azimuthal correlations $dN_{pair}/d\Delta\phi$ in central AuAu collisions
shows a ``dip'' with a ``double peak'' structure at $\Delta\phi\approx\pi\pm$ 1.1 (Fig.~\ref{fig:dNdphi}) 
for semihard associated hadrons ($p_{T,\,assoc}$ = 1 -- 2.5 GeV/$c$)~\cite{phenix_machcone,star_machcone}.
Such a non-Gaussian (``volcano''-like) profile in the away-side hemisphere has been interpreted as due to 
the preferential emission of energy from the quenched parton at a finite angle with respect to the jet axis.
Such a  conical-like pattern can appear if the lost energy excites a collective mode of the medium and
generates a wake of lower energy gluons with Mach-~\cite{mach,rupp05} or \v{C}erenkov-like~\cite{rupp05,cerenkov} 
angular emissions. In the first case, the {\it speed of sound}\footnote{The $c_s$ of an ultrarelativistic fluid 
is a simple proportionality constant relating its pressure and energy density: $P=c_s^2\;\varepsilon$.}
of the traversed matter, $c_s^2 = \partial P/\partial\varepsilon$, can be determined from the characteristic supersonic 
angle of the emitted secondaries: 
$\cos(\theta_{M}) = c_s$, where $\theta_{M}$ is the Mach shock wave angle.
The resulting preferential emission of secondary partons from the plasma measured 
at a {\it fixed} angle $\theta_{M} \approx 1.1$, 
yields an average value of the speed sound $c_{s}\approx$ 0.45, larger than that of a hadron gas 
($c_s \approx$ 0.35)~\cite{alam03}, and not far from that of an ideal QGP ($c_s = 1/\sqrt{3}$).


\section{Summary}
A selection of experimental data from central AuAu collisions at RHIC energies  ($\sqrtsnn$ = 200 GeV) 
has been presented providing direct information on fundamental thermodynamical and transport properties 
of high-density QCD matter. Four notable experimental results have been discussed: 
(i) the reduced total hadron multiplicities consistent with gluon saturation in the initial nuclear parton
distribution functions, (ii) the strong transverse and elliptic differential flows in the bulk hadron spectra 
indicative of a high degree of collectivity and very low viscosities during the first instants of the reaction, 
(iii) the factor $\sim$5 suppression of high $p_T$ leading hadrons reproduced by parton energy loss 
calculations for a medium with very large initial gluon densities ($dN^g/dy\approx$ 1000) and 
transport coefficient ($\mean{\hat{q}}\approx$ 14 GeV$^2$/fm), and 
(ii) the non-Gaussian shape of the azimuthal distributions of secondary hadrons in the away-side 
hemisphere of high-$p_T$ trigger hadrons attributed to Mach conical flow caused by the propagation of 
a supersonic parton through the dense system. Other interesting probes of quark-gluon matter (photons,
quarkonia, ...), not covered here by lack of space, will be discussed in a coming publication~\cite{dde06}.
Nucleus-nucleus collisions at LHC energies will undoubtedly contribute to expand the knowledge of
many-body QCD at extreme conditions of temperature, density and low-$x$ shedding light on a vast 
ramification of fundamental physics problems.\\



\noindent {\bf Acknowledgments}\\
\noindent This work is supported by the 6th EU Framework Programme contract MEIF-CT-2005-025073.


\end{document}